\documentclass[a4paper,aip,jcp,citeautoscript,reprint,nofootinbib]{revtex4-2}

\usepackage[T1]{fontenc}
\usepackage[utf8]{inputenc}

\usepackage{booktabs}
\usepackage{threeparttable}

\usepackage[normalem]{ulem}

\usepackage{graphicx}
\usepackage{hyperref}
\hypersetup{
  colorlinks,
  citecolor=blue,
  linkcolor=blue,
  urlcolor=blue
}
  
\usepackage{siunitx}
\DeclareSIUnit\bar{bar}
\DeclareSIUnit\angstrom{\text {Å}}
\DeclareSIUnit\bohr{\text {b}}

\usepackage{textgreek}
\usepackage{chemmacros}
\usepackage{amssymb}

\usepackage{float}

\newcommand{\ESI}[1]{\textcolor{blue}{{#1}}}

\begin{document}

\title{Accounting for the Quantum Capacitance of Graphite in Constant Potential Molecular Dynamics Simulations}

\author{Kateryna Goloviznina}
\affiliation{Sorbonne Universit\'e, CNRS, Physicochimie des \'Electrolytes et Nanosyst\`emes Interfaciaux, F-75005 Paris, France}
\affiliation{R\'eseau sur le Stockage Electrochimique de l'Energie (RS2E), FR CNRS 3459, 80039 Amiens Cedex, France}

\author{Johann Fleischhaker}
\affiliation{Institute of Polymers and Composites, Hamburg University of Technology, 21073 Hamburg, Germany}
\affiliation{Sorbonne Universit\'e, CNRS, Physicochimie des \'Electrolytes et Nanosyst\`emes Interfaciaux, F-75005 Paris, France}

\author{Tobias Binninger\footnotemark{}}
\affiliation{ICGM, Univ Montpellier, CNRS, ENSCM, Montpellier, France}
\altaffiliation[{Present} address: ]{Theory and Computation of Energy Materials (IEK-13), Institute of Energy and Climate Research, Forschungszentrum J\"ulich GmbH, 52425 J\"ulich, Germany}

\author{Benjamin Rotenberg}
\affiliation{Sorbonne Universit\'e, CNRS, Physicochimie des \'Electrolytes et Nanosyst\`emes Interfaciaux, F-75005 Paris, France}
\affiliation{R\'eseau sur le Stockage Electrochimique de l'Energie (RS2E), FR CNRS 3459, 80039 Amiens Cedex, France}

\author{Heigo Ers}
\affiliation{University of Tartu, Ravila 14a, 51004 Tartu, Estonia}

\author{Vladislav Ivanistsev}
\affiliation{University of Tartu, Ravila 14a, 51004 Tartu, Estonia}

\author{Robert Meissner}
\affiliation{Institute of Polymers and Composites, Hamburg University of Technology, 21073 Hamburg, Germany}
\affiliation{Institute of Surface Science, Helmholtz-Zentrum Hereon, 21502 Geesthacht, Germany}

\author{Alessandra Serva}
\affiliation{Sorbonne Universit\'e, CNRS, Physicochimie des \'Electrolytes et Nanosyst\`emes Interfaciaux, F-75005 Paris, France}
\affiliation{R\'eseau sur le Stockage Electrochimique de l'Energie (RS2E), FR CNRS 3459, 80039 Amiens Cedex, France}

\author{Mathieu Salanne}
\email{mathieu.salanne@sorbonne-universite.fr}
\affiliation{Sorbonne Universit\'e, CNRS, Physicochimie des \'Electrolytes et Nanosyst\`emes Interfaciaux, F-75005 Paris, France}
\affiliation{R\'eseau sur le Stockage Electrochimique de l'Energie (RS2E), FR CNRS 3459, 80039 Amiens Cedex, France}
\affiliation{Institut Universitaire de France (IUF), 75231 Paris, France}

\date{\today}


\begin{abstract}

Molecular dynamics simulations at a constant electric potential are an essential tool to study  electrochemical processes, providing microscopic information on the structural, thermodynamic, and dynamical properties. Despite the numerous advances in the simulation of electrodes, they fail to accurately represent the electronic structure of materials such as graphite. In this work, we introduce a simple parameterization method that allows to tune the metallicity of the electrode based on a quantum chemistry calculation of the density of states. As a first illustration, we study the interface between graphite electrodes and two different liquid electrolytes, an aqueous solution of NaCl and a pure ionic liquid, at different applied potentials. We show that the simulations reproduce qualitatively the experimentally-measured capacitance; in particular, they yield a minimum of capacitance at the point of zero charge, which is due to the quantum capacitance contribution. An analysis of the structure of the adsorbed liquids allows to understand why the ionic liquid displays a lower capacitance despite its large ionic concentration. In addition to its relevance for the important class of carbonaceous electrodes, this method can be applied to any electrode materials (e.g. 2D materials, conducting polymers, etc), thus enabling molecular simulation studies of complex electrochemical devices in the future.

\end{abstract}

\maketitle

\section{Introduction}

The electrochemical double-layer (EDL) is the region where the electrolyte contacts the electrode in an electrochemical device. Despite its importance in many fields, ranging from energy storage,~\cite{salanne2016a} electrocatalysis,~\cite{ledezmayanez2017a} sensing,~\cite{chipangura2023a} water desalination, nanogenerators~\cite{pace2023a}, \textit{etc.}, much remains to be known about its structure and properties. \textit{In situ} experiments provide many useful details,~\cite{forse2016a} but they are often hindered by their spatial resolution since the processes at play occur in a liquid region of a few nanometers only. From the theoretical point of view, although textbooks still present the Gouy--Chapman--Stern model as a realistic picture, its relevance for all these applications has largely been questioned due to its drastic approximations.~\cite{kornyshev2007a,wu2022a} Based on these observations, microscopic simulations appear as the ideal tool for advancing our knowledge about the EDL.~\cite{jeanmairet2022a}

Yet, the application of microscopic simulations to such complicated interfacial systems is not straightforward. Researchers are faced with the conundrum of simulating large-scale systems, over relatively long timescales, while keeping the accuracy of the calculation to the highest level. For example, the characteristic adsorption lifetimes of ionic species inside electrified nanoporous electrodes are of several nanoseconds.~\cite{pean2015b} Recent developments have made density functional theory (DFT) accessible to calculate the interactions in  molecular dynamics (MD) simulation, but the associated computational costs are very large.~\cite{gross2023a} This method was thus  so far limited to the important case of metal-water interfaces,~\cite{gross2022a} with a special focus on the reactivity of water molecules.~\cite{surendralal2018a} In recent years, several works have intended to account for the presence of ions by including them directly within the electrochemical double-layer. This approach yielded accurate estimates of the interfacial capacitance at the interfaces between gold,~\cite{li2019b} copper or graphene~\cite{li2023a} and simple aqueous solutions, but the limited system size prevents one from controlling the concentration of ions. A similar setup was used in the framework of electrocatalysis applications to simulate the effect of various cations on the CO$_2$ reduction reaction mechanism.~\cite{monteiro2021a,qin2023a} The main example of non-aqueous system simulated using DFT-based MD is the study of the potential drop at the interface formed between a pure ionic liquid and a graphene electrode,~\cite{Ers2020} but it is worth noting that the computational cost of the simulations was of the order of 1~million CPU hours to obtain trajectories of $\approx$1~ps. For applications in which the effects of ion adsorption are important and an efficient sampling of the liquid configurations needs to be performed, the use of less costly alternative is therefore necessary.

Consequently, the main method currently employed for simulating electrochemical double-layers is classical MD, in which the interactions are represented using a classical force field.~\cite{jeanmairet2022a} In current electrode models, the polarization by the electrolyte is generally accounted for by allowing the partial atomic charges of the electrode atoms to fluctuate over time, while maintaining a constant potential difference between the negative and the positive electrodes.~\cite{siepmann1995a,reed2007a} However, this approach did not include the so-called quantum capacitance (QC) effects. As it has been known since the pioneering work of Gerischer on graphite,~\cite{gerischer1985a} the density of states (DOS) of the electrode may vary near the Fermi level, which reflects on the electrode ability to respond to the polarization by the electrolyte. A simple solution to this problem was found by exploiting the fact that the EDL and quantum capacitance contributions are in series, and could thus be obtained by two different calculations.~\cite{kornyshev2013a,pak2013a} The QC can then be readily obtained at the DFT level from a static calculation; a recent example of this approach is the determination of the capacitance of graphdiynes, a new class of 2D porous materials that display a complex electronic structure.~\cite{mo2023b} While it gives accurate capacitances, this approach cannot yield a good structural description of the EDL at a given potential since the interaction of the electrolyte with the electrode is not modified accordingly in the computational workflow. Recently, several attempts to account for the metallicity of the electrode were reported, in particular in the framework of the Thomas--Fermi model.~\cite{scalfi2020b,schlaich2022a} The constant-potential version of this model includes an additional parameter, the Thomas--Fermi length, that accounts for the screening of the potential within the electrode. It was used to systematically study the effect of the electrode metallicity on  properties such as the interfacial free energies.~\cite{scalfi2021b} In principle, the Thomas--Fermi model includes intrinsically the QC effects, yet the use of a single parameter that is material-dependent does not allow to investigate the impact of the variations of the number of charge carriers in the electrode with the applied potential.

In this work, we show how to account explicitly for the DOS of an electrode material in classical molecular dynamics simulations. To this end, we selected graphite as a representative material due to its ubiquitous importance and to its peculiar DOS in the vicinity of the Fermi level. The parameterization requires a simple DFT calculation, and the model can be used in the current implementations of the constant potential method. We show its application by computing the full EDL capacitance, which now contains the quantum contribution, for two different electrolytes, aqueous NaCl (with a concentration of  1~M) and the 1-ethyl-3-methylimidazolium tetrafluoroborate ionic liquid (\ch{[EMIM][BF4]} IL). We show that the agreement with experiments is almost quantitative, in particular our simulations reproduce the minimum of capacitance at the point of zero charge (PZC). This then allows us to discuss the microscopic details of the systems; in particular we show that despite its much lower ionic concentration the aqueous system displays a larger capacitance due to a strong contribution from the water molecules. In the future, this approach will enable the study of electrode materials with more complex electronic structure, in particular in the context of capacitive energy storage.\cite{wang2017m,mo2023b,zhu2016a}

\section{Model}
\subsection{The electrode model and its parameterization}

\begin{figure*}[htb]
  \centering
  \includegraphics[width=17cm]{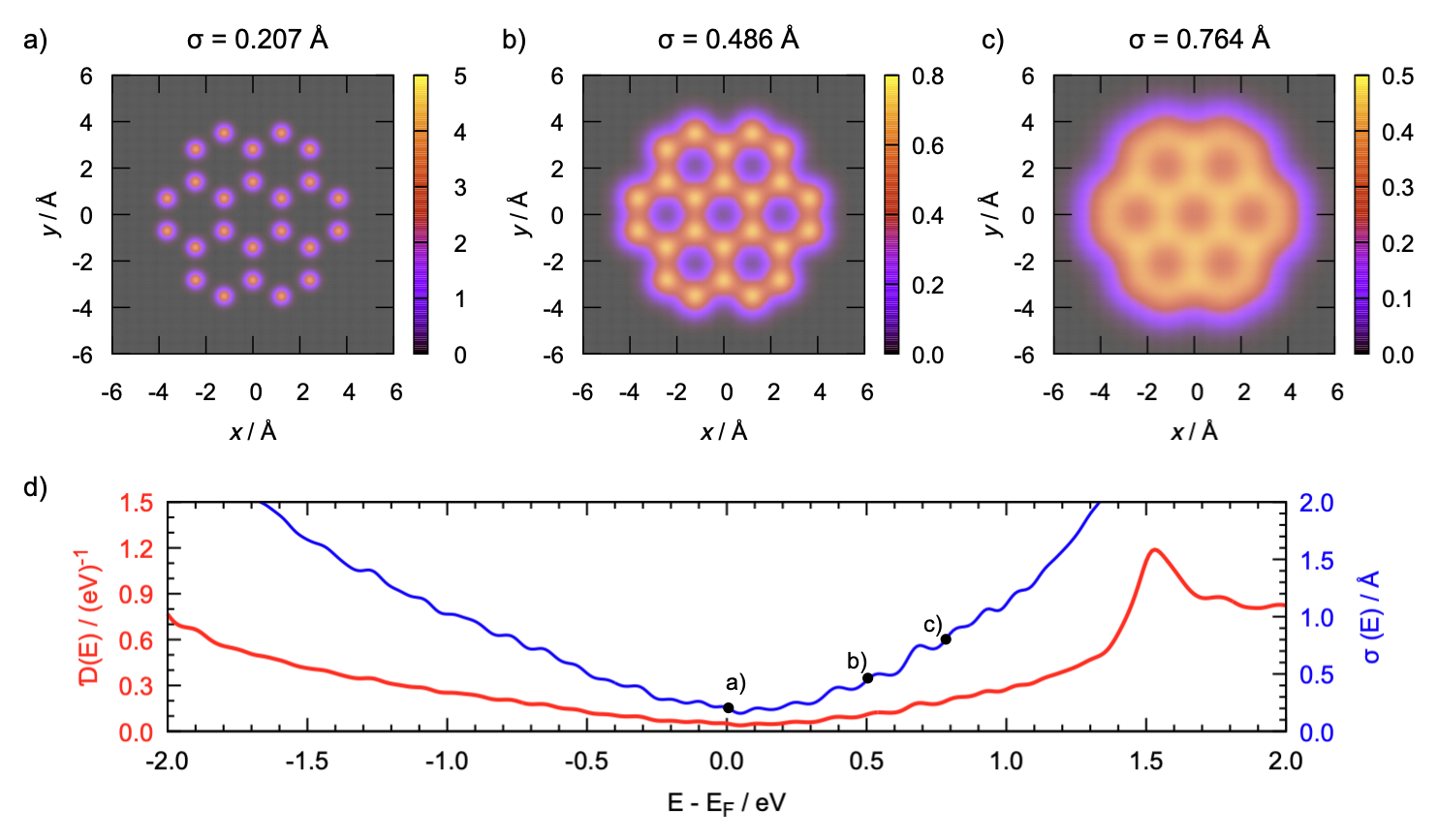}
  \caption{ (a-c) Illustration of the impact of the Gaussian width ($\sigma$) on the charge density of a coronene molecule. For visualization purposes, two-dimensional charge distributions centered on the carbon atoms were used in this plot. The values in the colorbar are given in $e$/\AA$^2$. The represented charge density corresponds to a case where all the partial charges of the carbon atoms are equal; in the constant potential model, they fluctuate in response to the electrolyte. (d) Density of states of graphite obtained from DFT calculations}
  \label{fig:dos} 
\end{figure*}

The constant potential electrode model introduced by Siepmann and Sprik~\cite{siepmann1995a} is based on a representation of the electrode charge distribution, $\rho_{\rm elec}$, as a collection of atom-centered Gaussians:

\begin{equation}
	\rho_{\rm elec} (\textbf{r}) = \sum_{i=1}^{N_{\rm elec}} q_i \dfrac{1}{(2 \pi \sigma^2)^{3/2}} \exp \Bigg (-\dfrac{\mid\textbf{r} - \textbf{r}_i\mid^2}{2 \sigma^2} \Bigg ),
\end{equation}	

\noindent where $N_{\rm elec}$ is the number of atoms (with positions ${\textbf{r}_i}$) in the electrodes and $\sigma$ is the width of the Gaussian. The partial charges $q_i$ of the electrode atoms behave as additional degrees of freedom of the simulation, contrary to the electrolyte charges represented using point charges of fixed magnitude. The fluctuating charges should take values such that the potential is uniform within the whole electrode at each timestep. In practice, we define two electrodes with potentials $\Psi_+$ and $\Psi_-$ (and with the corresponding number of atoms $N_+$ and $N_-$). The determination of the fluctuating charges is done within the Born--Oppenheimer approximation, by minimizing the total electrostatic potential energy of the system which includes all the Coulomb interactions, as well as the electric work which is necessary to charge the electrodes to the desired potential. The corresponding expressions are provided in Reference \citenum{coretti2022a}.  

In the original model, this potential energy term includes only one parameter, which is the width of the Gaussian (taken the same for all the electrode atoms). Tuning its value has two effects. Firstly, larger Gaussians lead to an increased delocalization of the electrode charge. This effect can be appreciated in the example of two-dimensional charge distributions for a coronene molecule in Figure \ref{fig:dos}a-c, in the specific case for which all the atoms would carry the same partial charge. For example, for $\sigma=\SI{0.207}{\angstrom}$, it is highly localized around the atoms. Some overlap starts to be seen for a value of \SI{0.486}{\angstrom}, as a non-negligible charge density starts to be observed along the C-C bonds. Finally, for a larger value of \SI{0.764}{\angstrom}, the charge density is now almost equally distributed within the carbon rings. The latter representation is in agreement with pictures extracted from single-molecule imaging of aromatic carbonaceous molecules using scanning tunneling microscopy or atomic force microscopy.~\cite{gross2009a} Secondly, larger Gaussian widths result in an enhanced ability to charge the atom. Indeed, the Gaussian charges self-interact through the term:

\begin{equation}
U_{\rm self}= \frac{1}{4\pi\epsilon_0}\left(\frac{1}{\sigma\sqrt{4\pi}}  \sum_{i \in (N_++N_-)}q_i^2\right),
\end{equation}

\noindent \noindent so that the magnitude of this self-interaction corresponds to an energy penalty for charging the electrode. 

The extension of the model to include Thomas--Fermi effects simply introduced an additional energy penalty proportional to the Thomas--Fermi screening length $l_{\rm TF}$:
\begin{equation}
U_{\rm TF}=\frac{1}{4\pi\epsilon_0}\left(2\pi l_{\rm TF}^2 \rho \sum_{i \in (N_++N_-)}q_i^2\right) \label{eq:thomasfermienergy}
\end{equation}
\noindent where $\rho$ is the number density of electrode atoms within the material. Note that this term does not impact the interactions between neighbor atoms and the delocalization of the charge within the plane, which remained controlled by the Gaussian width. Quantities such as the  interfacial capacitance or the solid-liquid interfacial free energies were shown to vary systematically with $l_{\rm TF}$ in previous works.~\cite{scalfi2020b,scalfi2021b}

Here our objective is to go further and use the DOS of the material ($\mathcal{D}(E)$) to parameterize the electrode model. The DOS is defined as a measure of the number of allowed electronic states whose energies lie at a value $E$. A DOS plot of graphite, calculated at the DFT level (details given in the SI), is shown in Figure \ref{fig:dos}d. It displays a characteristic minimum around the Fermi level $E_\mathrm{F}$, which often leads graphite to be classified as a ``semi-metal''. To accurately describe the quantum capacitance effect, the residual value of the DOS at the Fermi level must be correctly quantified, requiring the use of a very dense $K$-point mesh and a consistent treatment of thermal broadening in the calculation of the DOS (see SI for details) \cite{Binninger2021}. The Thomas--Fermi length can be expressed in terms of the DOS at the Fermi level:
$l_{\rm TF} = \Bigg (\dfrac{e^2}{\epsilon_0} \mathcal{D}(E_F) \Bigg)^{-1/2}$
\noindent which provides an easy parameterization of the model through Equation \ref{eq:thomasfermienergy}. This approach was successfully used in a theoretical model to explain the difference in freezing properties behavior for an ionic liquid in contact with electrodes of varying metallicities.~\cite{comtet2017a}

Here we extend this approach by using the fact that in the fixed band approximation, the Fermi level is shifted by the bias potential to $E_F+e\Psi$, which leads to an increase in the number of available states. We use this correspondence to determine the DOS at a given electric potential, which in turn gives a new value for the Thomas--Fermi screening length. However, a difficulty arises in choosing the bias potential since in the constant potential electrode model; the fixed quantity is the applied potential ${\Delta\Psi}$ between the electrodes. A constraint of global electroneutrality is imposed,~\cite{scalfi2020a} so that for ${\Delta\Psi}=0$ the two electrodes are equivalent. Their charge must fluctuate around zero, and we can define their potential as the PZC ($\Psi_+=\Psi_-=\Psi_{\rm PZC}$). For non-null applied voltages, there exists a potential shift that depends on the instantaneous configurations and fluctuates around an average value, so that $\Psi_+=\frac{\Delta\Psi}{2}+\Psi_{\rm shift}$ and $\Psi_-=-\frac{\Delta\Psi}{2}+\Psi_{\rm shift}$. In the following, based on the symmetric shape of the experimental capacitance, we arbitrarily neglect the variation of the electrode potentials ($\Psi_{\rm shift}=0$\,V, but in future work, it should be possible to allow the Thomas--Fermi length to vary dynamically to account for it).

In practice, we chose to use the Gaussian width as the main parameter to control the DOS of the simulation instead of the $l_{\rm TF}$. This leads to the following parameterization for $\sigma$ depending on the potential:
\begin{equation}
\sigma(\Psi) = \frac{1}{4\pi \epsilon_0}\left(\frac{e^2\mathcal{D}(E_F+e\Psi)}{\rho\sqrt{\pi}}\right)
\end{equation}
The obtained values of Gaussian width are given in Supplementary Table~\ESI{S1}. 
At the PZC, the Gaussian width takes a minimal value of 0.207\,\AA\, which corresponds to the situation shown in Figure~\ref{fig:dos}a for coronene.
Since the DOS slightly differs depending on the sign of the potential, we considered non-identical Gaussian width for the negative and positive electrodes. However, for the largest applied potential of 2\,V, the corresponding Gaussian width would be larger than 1.0\,\AA. This value corresponds to a situation where the overlap between neighboring Gaussians is too large, and the problem becomes ill-defined mathematically.~\cite{gingrich2010a} In practice, we observed that the chosen $\sigma$ cannot be larger than \SI{0.764}{\angstrom}, which corresponds to the case shown on Figure~\ref{fig:dos}c for coronene. This value was thus selected for the 2\,V potential simulation for both the positive and negative electrode atoms.

\subsection{Non-Coulombic electrode/electrolyte interactions}

Additional parameters need also to be defined on the electrolyte side. 
Although, for the aqueous solution, a well-established non-polarizable force field is employed, it is not the case for the \ch{[EMIM][BF4]} ionic liquid, for which we use a polarizable interaction potential. The induced dipoles are calculated at each timestep together with the fluctuating charges, using a preconditioned conjugate gradient method as described in Reference \citenum{Coretti2022}. In such a case, the parameters of the conventional OPLS force field cannot be used in order to avoid double-counting of some interactions as explained briefly in the following. We therefore use the CL\&Pol approach,\cite{Padua2017,Goloviznina2019} which consists in scaling down the non-bonded attractive term (in the case of the Lennard Jones potential, the $\epsilon$ parameter) by the factor $k_{ij}$:
\begin{equation}
  \label{eq:kij}
    k_{ij} = \frac{E_{\mathrm{disp}}}{E_{\mathrm{disp}} + E_{\mathrm{ind}}}  
\end{equation}
so that the potential accounts only for dispersion interactions, while the implicit induction contribution is removed. The scaling factor can be evaluated from the symmetry-adapted perturbation theory (SAPT) calculation, which allows isolating the dispersion and induction terms from the total interaction energy of two fragments. IL ions are treated as entire fragments, while a coronene molecule is considered as representative of the carbon nanomaterial. The geometries of corresponding dimers (cation--coronene, anion--coronene, and coronene--coronene) are optimized, and then potential energy scans are performed at different SAPT levels (details given in SI). An example of the SAPT scan is given in Supplementary Figure~\ESI{S2}, and the final $k_{ij}$ values are reported in Supplementary Table~\ESI{S2}.

\begin{figure}[htb]
  \centering
 \includegraphics[width=8.5cm]{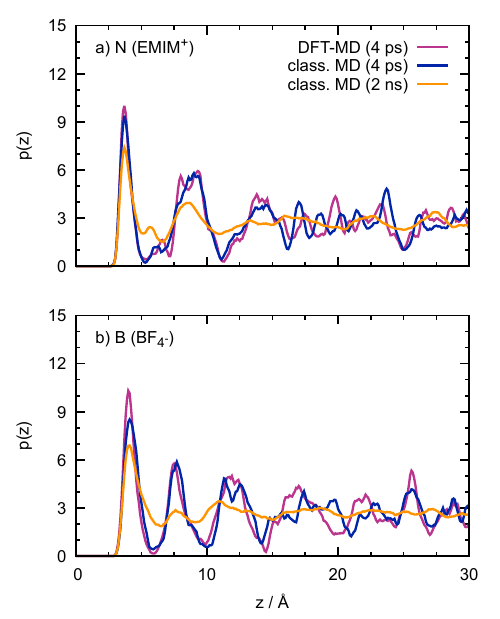}
  \caption{Normalised density profiles of selected atoms of \ch{[EMIM][BF4]} as a function of their distance from the electrode. Rose and blue lines correspond to the trajectories of the same length, using the same initial configuration ($\sim$~\SI{4}{\pico\second}), while the orange line corresponds to a long \SI{2}{\nano\second} run.}
  \label{fig:zdf_val} 
\end{figure}

In order to validate the proposed model, we compared the atomic forces (given in Supplementary Figure~\ESI{S3}) and density profiles to the ones of a DFT-based MD simulation. We re-used a model (Ref. \citenum{Ers2020}), which represents a single graphite electrode with $3$ layers and $200$ ion pairs of \ch{[EMIM][BF4]}. As discussed above, due to the high computational cost associated with the DFT-MD method (see methods description in SI), the total simulation time was only around 5\,ps. Nevertheless, it allowed us to check whether the equilibrium distances between the ions and the electrodes are well reproduced. An example of density profiles is given in Figure~\ref{fig:zdf_val}. Both in the cases of the N atom of the \ch{EMIM+} cation and the B atoms of the \ch{BF4-} anion, we observe an excellent agreement between two methods in the peak positions and intensities up to $\sim$\SI{15}{\angstrom} from the surface. The positions of the two first cationic and anionic layers are also preserved when a long \SI{2}{\nano\second} run is carried out. The intensities of these peaks then decrease as expected from the better sampling of the configuration space. 

\subsection{Results and discussion}

As a first application of our method, we focus on the canonical example of the graphite electrode. By subtracting the EDL capacitance (estimated using analytical contributions associated with the Helmholtz and the diffuse layer) from the experimentally measured differential capacitance, in 1985, Gerischer determined the QC, and hence the DOS around the Fermi level, in qualitative agreement with the theoretical calculations available at the time.~\cite{gerischer1985a}  Here we use an inverse approach, in which the DOS is directly fed into the model, with the aim to determine the differential capacitance without any empirical input (other than the one used to build the force fields).

When using the constant potential method, a direct output of the simulation is the total charge ($Q_{\rm elec} = \sum_{i \in N_{\rm elec}} q_i$) on the electrodes. An electroneutrality constraint is imposed so that $Q_+=-Q_-=Q$ at each timestep. In order to determine the capacitance, it is necessary to extract the average value of $Q$ once the equilibrium has been reached at a given potential. As shown in Supplementary Figure~\ESI{S4}, this regime is obtained well before a time of 1\,nanosecond of simulation for the aqueous solution, but several nanoseconds are necessary in the case of the ionic liquid due to its much larger viscosity. 

\begin{figure}[htb]
  \centering
 \includegraphics[width=8.5cm]{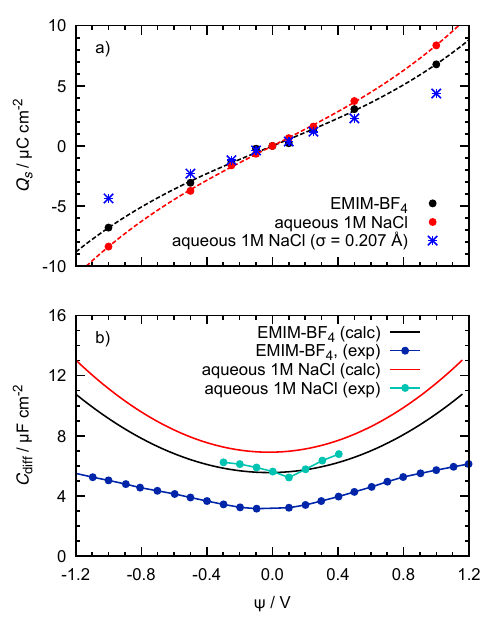}\\
 \caption{(a) Average surface charge on the graphite with respect to the electrode potential (symbol: simulation data, dashed line: fit with a third-order polynomial function). The blue stars are the surface charges obtained using a constant Gaussian width of 0.207~\AA, \textit{i.e.}, the Gaussian value obtained from the DOS at \SI{0}{V}. (b) Differential capacitance dependence on the electrode potential. The experimental values are extracted from references \citenum{Oll2017} and \citenum{Iamprasertkun2019}.}
  \label{fig:capa} 
\end{figure}



The differential capacitance is then calculated by deriving the variation of $<Q>$ with respect to the electrode potential, $\Psi_+$ or $\Psi_-$.
In practice, due to the limited number of points, a fit of the $Q=f(\Psi)$ plot is necessary prior to the derivation. As can be seen in Figure \ref{fig:capa}a, a third-order polynomial function yields a very good fit of the data.   Comparison with the experimental results obtained for the same electrolytes~\cite{Oll2017,Iamprasertkun2019}, shown in Figure \ref{fig:capa}b as dotted lines, gives qualitative agreement. In particular, our simulations reproduce the minimum of capacitance at the point of zero charge (which corresponds by construction to an applied potential of 0\,V in our case). Still, they overestimate the value of this minimum by approximately 2\,{\textmu}F\,cm$^{-2}$. This discrepancy may be attributed to either some errors due to the employed force field or to the fact that we simulate a perfect, pristine graphite surface while the experimental materials may not be perfectly planar despite the use of highly ordered pyrolytic graphite. Despite this difference, the variation of the differential capacitance with potential is very similar between simulations and experiments, which shows that the response to applied voltage is well taken into account. In order to show further the impact of the QC in the simulation, additional simulations of the aqueous NaCl solution were performed with a fixed Gaussian width of 0.207\,{\AA} (that corresponds to the value taken at rest in our scheme). The accumulated charge is then much smaller, as can be seen in Figure \ref{fig:capa}a (blue stars), and its variation is better fitted with a linear function, which corresponds to a constant interfacial capacitance.

\begin{figure}[t!]
  \centering
 \includegraphics[width=8.5cm]{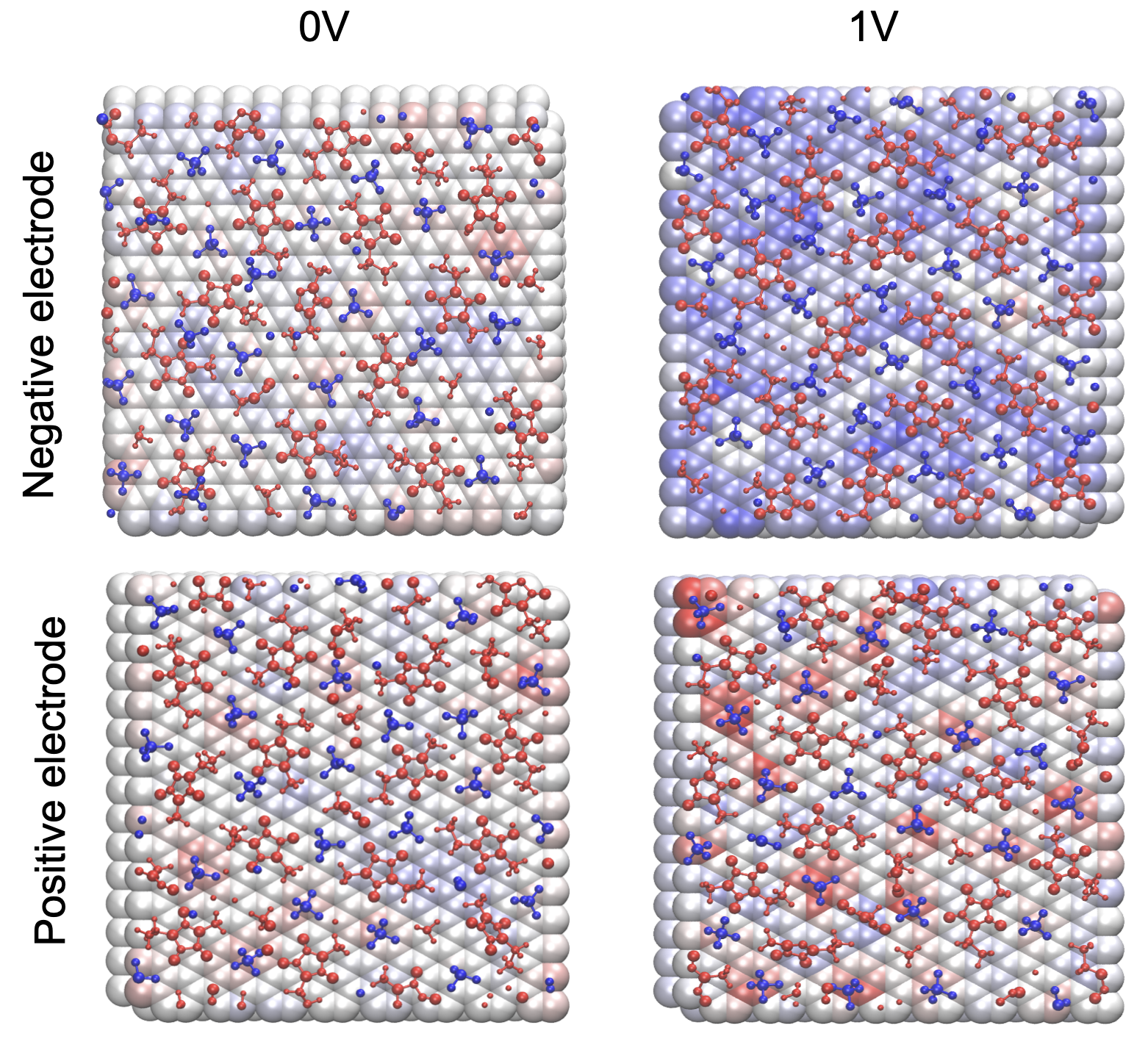}
  \caption{Representative snapshots of the adsorbed layer for \ch{[EMIM][BF4]} IL-based simulations: \ch{EMIM^+} is given is red, \ch{BF4-} in blue. The surface atoms are colored according to their charge (white: neutral, blue: negative, red: positive). }
  \label{fig:adsorbed_IL} 
\end{figure}

Notably, as in experiments, when accounting for the QC we observe that the capacitance of the aqueous solution is larger than the one of the ionic liquid. This result may be surprising since the latter is entirely composed of ions, and could thus be expected to induce a larger charge at the surface of the electrode. It is, therefore, interesting to compare the local structure of the two systems and try to understand the microscopic mechanism at play. The structure of the two interfaces was largely documented in various simulation studies,\cite{Haskins2016,Ferreira2022,He2024,Hu2013} but in the absence of explicit QC effect and no systematic comparison was performed so far. On the one hand, in the ionic liquid, at zero applied electric potential, a ``checkerboard-like'' structure is obtained in which the first liquid layer is made of well-organized adsorbed ions. In particular, as shown in Figure~\ref{fig:adsorbed_IL}, the imidazolium rings tend to lie flat on the surface\cite{Carstens2014} and to arrange collectively along lines. This is consistent with previous observations based on atomic force microscopy experiments as well as previous MD simulations on similar systems.~\cite{hayes2010a,black2013b} Upon application of a voltage, due to steric hindrance, the charge can only accumulate \textit{via} an exchange mechanism. In the case of the negative electrode and for an applied voltage of \SI{2}{V}, the cations also tend to reorient and become perpendicular to the electrode (Supplementary Figure~\ESI{S5}), thus allowing a larger ionic density on the surface. 

\begin{figure}[htb]
  \centering
 \includegraphics[width=8.5cm]{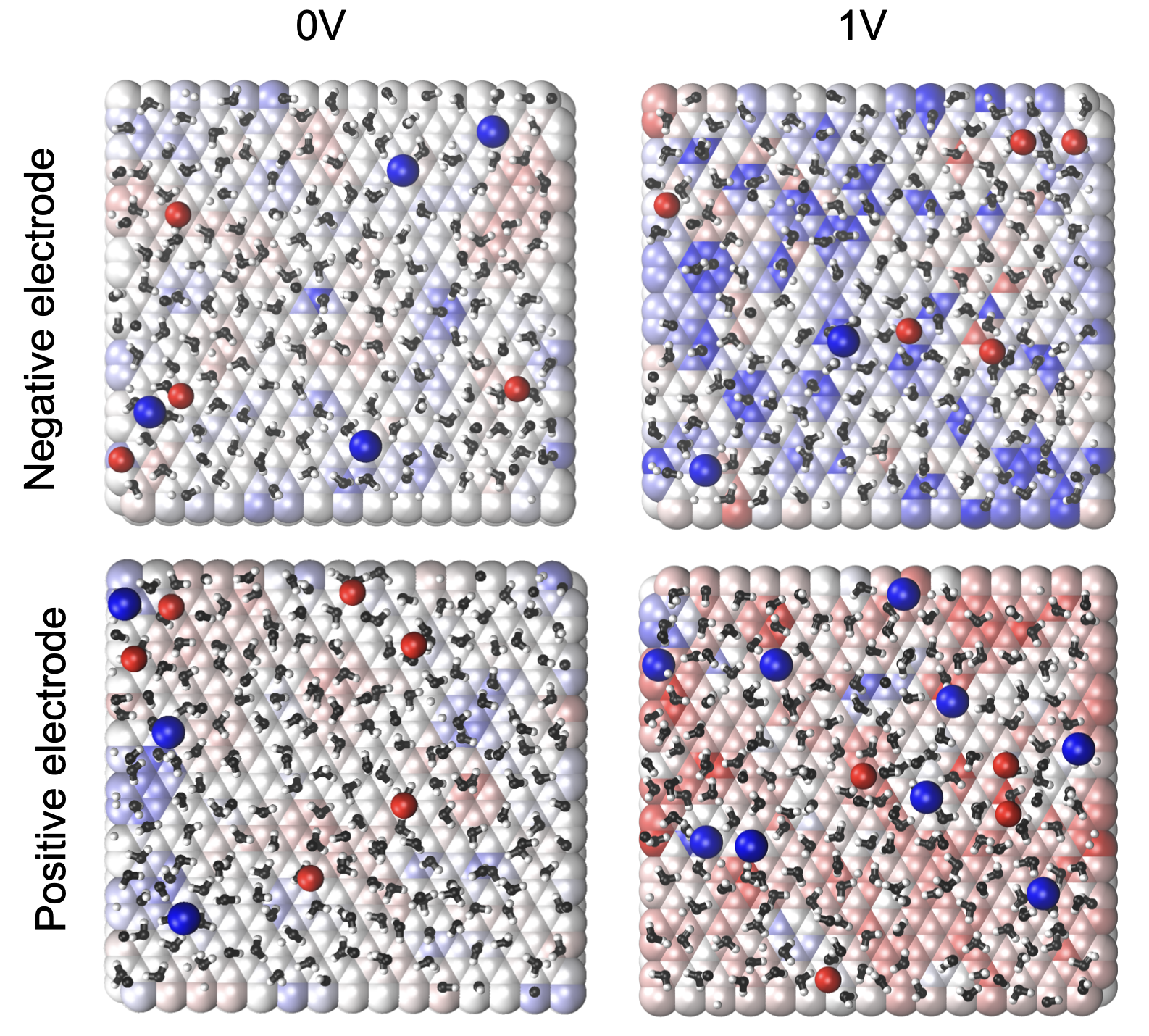}\\
  \caption{Representative snapshots of the adsorbed layer in the case of \SI{1}{M} aqueous NaCl (black and white: water molecules, red spheres: Na$^+$ ions, blue sphere: Cl$^-$ ions). The surface atoms are colored according to their charge (white: neutral, red: positive, blue: negative). The represented Na$^+$ and Cl$^-$ ions are located beyond the plane of water molecules, and they do not induce any noticeable surface charge on the electrode.}
  \label{fig:adsorbedwater} 
\end{figure}

On the contrary, in the case of the aqueous sodium chloride solution, we do not observe any adsorption of the Na$^+$ and Cl$^-$ in the first liquid layer at low potential (see Supplementary Figure~\ESI{S6}). Some ions are present in the second layer, but they do not lead to any noticeable surface charge on the electrode surface, as shown in Figure \ref{fig:adsorbedwater}. This is consistent with previous simulations performed on gold surfaces, in which it was shown that the degree of metallicity of the gold had to be very high to promote their adsorption.~\cite{Serva2021} Similarly, no specific adsorption of  Na$^+$ and Cl$^-$ ions was observed in a previous work when the Thomas--Fermi length was systematically varying.\cite{scalfi2020b} The polarization of the surface is thus almost entirely due to the behavior of the liquid water molecules. At zero voltage, they adopt a parallel orientation with respect to the electrode surface, and they form a 2-dimensional hydrogen bond network inside the plane. Up to one volt, the charging of the electrode does not imply any variation on the composition of the EDL, but it is rather accompanied by a change of the orientation of the water molecules, as shown in Supplementary Figure~\ESI{S7}. In line with our results, a recent simulation study of aqueous NaCl at different concentrations between Pt electrodes has shown that the electrostatic potential of the EDL is almost independent on the ionic concentration due to the molecular influence of water molecules.\cite{limaye2024}

Based on the different charging mechanisms, it is now possible to interpret the differences in capacitance between the two systems. From the energy point of view, charging the ionic liquid EDL requires breaking some strong Coulombic interactions between ions of opposite signs, while in the aqueous NaCl case, it only needs to rearrange the hydrogen bond network, which is substantially easier. Interestingly, this interpretation holds in the case of a planar surface, but it cannot be applied to more complex cases such as porous carbons for which the confinement effects lead to (i) an increased screening of the Coulombic interaction, which was termed as a ``superionic'' effect by Kondrat and Kornyshev~\cite{kondrat2011a,kondrat2023a} and confirmed by {\it in situ} X-ray diffraction experiments;~\cite{futamura2017a} and (ii) the impossibility to induce a macroscopic polarization through the water dipoles only, for symmetry reasons. Consequently, in the latter case, the polarization occurs \textit{via} the adsorption/desorption of the ions, and the capacitances are not larger anymore than for ionic liquids.~\cite{simoncelli2018a}

It is worth noting that the mechanisms described above are valid over the whole investigated potential range, and so there is interference from the QC on the overall behavior: It changes the onset potential at which the ion exchange/water reorientations occur. This validates the approaches based on a decoupling of the EDL and QC effect proposed in previous works,~\cite{kornyshev2013a,pak2013a,mo2023b} but our approach has the advantage of specifying the liquid structure at each potential without any ambiguity.

\section{Conclusions \& perspectives}

In conclusion, in this work we have shown how the constant potential method can be extended to explicitly account for the DOS of the material in MD simulations of electrochemical double layers. This extension can readily be used in any existing implementation, since it only requires fine-tuning of the Gaussian width of the electrode charge depending on the applied potential. The parameterization is easily made from a calculation of the DOS of the material at rest, using first-principles methods. Our new method, therefore, accounts directly for the QC contribution to the overall capacitance, contrarily to previous works in which it was simply added to the double-layer contribution. The advantage is that it is possible to extract the correct structure of the liquid at the interface for a given applied potential.

As a first application of the method, we focused on the graphite electrode. Despite its importance in energy applications and its apparent simplicity, its peculiar electronic structure is characterized by a minimum of the DOS at the Fermi level. Consequently, a correct description of the double-layer forming at its interface with electrolytes remained lacking so far. In this work, we obtained a qualitative agreement with experimental works for two important electrolytes, an aqueous NaCl solution and a pure ionic liquid. The simulations predict overall too large capacitances (by 1 to 2\,{\textmu}F\,cm$^{-2}$) for the two systems, but the variation with potential is well reproduced in both cases. The larger capacitance obtained for the aqueous system is interpreted based on the structure of the adsorbed liquid and its effect on the surface-induced charges.

In the future, this work could expand the scope of constant potential MD simulations to a variety of new materials for energy applications. In particular, many porous materials, such as carbon-based materials or metal-organic-frameworks, covalent organic frameworks, conducting polymers,~\cite{wang2017m} etc, display a complex electronic structure, that cannot be approximated to the one of a perfect conductor, and could benefit from such a representation. Additional work is however required to discriminate the presence of different chemical elements (with eventually different environments) in such materials. This could be tackled by coupling the present model with machine-learning-based approaches such as the PiNNwall interface~\cite{dufils2023a}, which was recently applied to the cases of chemically doped graphene and graphene oxide electrodes.

\begin{acknowledgments}
This work was funded by the European Research Council (ERC) under the European Union Horizon 2020 research and innovation program (grant agreements 771294 and 863473). It was also supported by the Estonian–French cooperation program Parrot, funded by the Estonian Research Council and Campus France Computational tasks were performed using resources from GENCI-IDRIS (Grant A0140910463). DFT-based MD simulations were run at the HPC resources provided by the Partnership for Advanced Computing in Europe (PRACE).
\end{acknowledgments}


\section{References}

%
\end{document}